# Practical Framework for Problem-Based Learning in an Introductory Circuit Analysis Course

Sebastian Martin
*Dept. of Electrical Engineering*
*Universidad de Malaga*
Malaga, Spain
smartin@uma.es

Salvador Pineda
*Dept. of Electrical Engineering*
*Universidad de Malaga*
Malaga, Spain
spineda@uma.es

Juan Perez-Ruiz
*Dept. of Electrical Engineering*
*Universidad de Malaga*
Malaga, Spain
jperez@uma.es

Natalia Alguacil
*Dept. of Electrical Engineering*
*Universidad de Castilla-La Mancha*
Ciudad Real, Spain
Natalia.Alguacil@uclm.es

Antonio Ruiz-Gonzalez
*Dept. of Electrical Engineering*
*Universidad de Malaga*
Malaga, Spain
afruiz@uma.es

*ABSTRACT*—Introductory courses on electric circuits at undergraduate level are usually presented in quite abstract terms, with questions and problems quite far from practical problems. This causes the students have difficulties to apply that theory to solve practical technical problems. On the other hand, electric circuits are everywhere in our lives, so we have plenty of real practical problems. Here we compile a selection of practical contexts suited for implementing Problem Based Learning approach in an introductory course on circuit analysis. And some examples describing the gamification process that uses these problems to build single-player role-playing games that fulfil the course contents and scheduling. The key point of the assessment and how it is related to the progress in the game is also described.

*INDEX TERMS*—Problem-Based Learning, situated learning, learning outcomes, role-playing game, assessment

## I. INTRODUCTION

The traditional approach for the introductory course on electric circuits at undergraduate level consists of using a quite abstract presentation of concepts, where the circuits to be solved are directly provided to the student without any connection to real technical problems. In this approach the usual complaints from the students are: i) problems are quite repetitive, what causes boredom, ii) students have difficulties in identifying the possible application of what they are learn- ing, it leads to a lack of motivation.

To solve these problems, lack of motivation and lack of connection with real practical problems, many approaches have been described in the literature. A strategy consisting of Problem-Based Learning (PBL) with situated learning (real context) implemented using gamification seems to be particu- larly appropriate.

These approaches have been extensively studied in the liter- ature, as for instance in [1] where a practical implementation of PBL for an elementary circuit analysis course is described, they propose to organize the class in working groups that have to solve open-ended problems. A similar approach is described in [2]. This approach is hard to apply if the number of students is high. To overcome this problem, computer aided implementations have been proposed in the literature, as for instance in [3] where problems are automatically generated and corrected, and students received feedback directly from the computer program. In this case the program generates many different problems, but they are not situated in a practical context (still quite abstract) and not open-ended.

Another interesting point to promote students engagement is implementing PBL using gamification, as is described for instance in [4]. One of the main difficulties is that the game must be properly related to the course topics.

Other difficulties with the application of these approaches are: i) the available time and workload for both the teachers and the students, particularly for the teachers, since the elabo- ration of the course material (problems, games, etc..) and the provision of feedback for each activity consume a lot of time, and ii) all the topics in the course program must be covered.

Here we propose an approach to overcome these difficulties based on: i) problems settled in real contexts, mainly equiva- lent circuits of real physical systems and of practical interest, ii) those problems are used as a part of classical role-playing games (paper and pencil). At the same time, the problems are suited to be automatically generated and corrected, but we do not deal with the automatic problem generation in this paper.

The main contributions related to the proposed conceptual framework are:

1) To add a real context to the problems (situated learning).
2) To link the elements involved in the problems (such as modeling physical systems as electrical circuits) and learning outcomes.
3) To create learning-training dynamics in the form of single-player role-playing games, selecting the sequence of problems to be solved and the elements involved in those problems.

The rest of the paper is organized as follows. The general Methodology is described in Section II. A selection of real contexts and equivalent circuits that can be used to define the problems in the course is listed on Section III. The link between the type of problems and the learning outcomes is discussed in Section IV.

Examples of role-playing games for an introductory course on circuit analysis are described in Section V. A summary of the main points in Section VI closes the paper.

## II. METHODOLOGY

The methodology developed in this paper consists of three main steps:

1) To elaborate a set of problems that correspond to equiva- lent circuits of real physical systems of practical interest (PBL in situated learning). These problems are described in Section III, and a list of sources with equivalent circuits that can be directly used in the problems is provided in Table I.

2) To relate the actions required to solve each problem to the learning outcomes (Bloom's taxonomy [5]). The link between learning outcomes and the proposed problems is discussed in Section IV.

3) Definition of the role-playing games. As the problems have a practical meaning, they can be grouped in se- quences to tell a story, and that story can be put in form of a role-playing game. Four examples of role-playing games are described in Section V.

Role-playing games described in Section V situate the student in four different context: i) a technician repairing domestic appliances, ii) an engineer designing the electric circuits in a power factory, iii) a consultant designing and checking the electric circuits for power supply in households, iv) a technician solving some circuit problems in a spaceship and a colony in Mars. Also, many other context can be defined, as for instance those related to: electric traction (subway, trains), electric installations in big buildings (malls, big office building) and so on.

The gamification process is described through examples in Section V, using the problems defined in Section III, and taking into account the objectives in terms of learning out- comes as defined in the course program. Here the gamification process is defined based on two main points:

4) Selection of contents: what defines the context, the kind of problems, and the link to the learning outcomes.

5) Assessment and the progress in the game. We follow an approach similar to that implemented in the classical role-playing games, with a main "quest" (main learning outcomes in the course) and some secondary quests. Setting levels, experience points and coins rewarded by solving each quest, the mechanism is described in detail in Section V. The experience points, level and skills in the game are interpreted in terms of learning outcomes.

This approach can be implemented within an admissible workload for the teacher, and also respecting the course pro- gram. It can be applied to courses with many students because the problems can be automatically generated and corrected, but we do not deal with the automatic problem generation in this paper. This framework contributes to both the students' engagement and the reaching of learning outcomes, because the students will learn the theory in the practical context and will be more confident to apply it in real problems.

## III. PROBLEMS IN REAL-WORLD CONTEXTS

Electrical circuits are everywhere in our lives: buildings, vehicles, mobile devices, domestic appliances, on the streets (lighting, distribution lines), and so on. Despite this, it results paradoxical that the contents in a first course on electrical circuits are usually presented in a very abstract form, without referring to real applications.

In this section a small sample of contexts and practical applications is described. Sources for equivalent circuits that can be directly used in an introductory course on circuit analysis are listed on Table I. Those equivalent circuits can be used to pose problems with a practical meaning and a real context. Some of them are related to advanced concepts not covered in an introductory course, but the problems can be posed to use just the information and the concepts in the introductory course. Also, we hope that the use of these contexts foster the students' curiosity, and animate them to enroll in other more advanced courses on related topics such as: electric machines, power generation, power transmission, power electronics, or smart grids.

In what follows the problems involved in the role-playing games are briefly described. These are typical practical prob- lems on electric circuits than can be posed as problems for an introductory course, since they properly fit in the usual course content:

· Checking the electric circuit in domestic appliances, in particular on that related to typical problems: short- circuit, open circuit, bypass currents, changing the topol- ogy (series, parallel), pure resistive circuits, Direct Cur- rent (DC), impedance, Alternating Current (AC), fre- quency response, power factor, improvement of the power factor.

· Wires routing and sizing in different contexts: a power plant, a household, a spaceship or a colony in Mars. We have to check voltage drops and power losses. It can be posed in different ways, for instance: resistors in DC circuits, or impedances in AC circuits. Note that each component in the circuits has a practical interpretation. Here it is important to highlight that loads in the systems usually are not constant and can take values in a range.

· Placement and calibration of protections in circuits. In- tegration of the loop to ground in the circuits. Study of the different failure modes in the circuits.

· Modification of the circuit topology to improve the reliability against possible failures, for instance includ- ing auxiliary circuits or additional branches to be con- nected/disconnected in case of failure.

· Improvement of the power factor by adding some com- ponents to reduce voltage drops and power losses.

· Integration of Photo Voltaic (PV) generation systems, with batteries considering their equivalent circuits. Tun- ing the parameters in the equivalent circuit for the max- imum power point tracking of the PV installation.

· Study of transformers and synchronous generators through their equivalent circuits, and their interaction with other components using three-phase circuits.

• Study of three-phase circuits present in neighborhoods, power plants, and so on.

A more detailed description of these problems for each game is provided in Section V.

## IV. LINK BETWEEN PROBLEMS AND LEARNING OUTCOMES

In this section we discuss the link between the proposed contexts and the learning outcomes. Instead of considering the traditional learning outcomes [5], the discussion is focused on the differences of the proposed approach with the traditional abstract presentation of the introductory course. First the typical contents in a introductory circuit analysis course are listed, then the main distinctive characteristics of the proposed approach are described. These characteristics are grouped in: i) real values and practical problems, ii) experiments and measures, and iii) references for additional content and self- learning.

### A. Contents in the introductory circuit analysis course

Contents in the introductory course are grouped in four blocks that follows:

• Basic concepts: loop, mesh, Kirchoff's laws, power and energy, ideal elements (resistor, capacitor and coil), in-dependent sources, parallel and series configurations, active and passive components, power balance, dependent sources, coupled coils, ideal transformers.
• Methods for circuit analysis: series connection and volt- age dividers, parallel connection and current divider, star- triangle transformation, real power sources and max- imum power transfer theorem, transformation of real sources. Mesh current analysis, nodal voltage analysis, analysis with dependent sources, linearity and superposi- tion, Tellegen's theorem, reciprocity theorem, Helmholtz- Thevenin's theorem, Helmholtz-Norton's theorem.
• Alternating current circuits: sinusoidal signals, phasors and phasor diagrams, response of basic elements (resistor, capacitor, coil), impedance, parallel and series connection of impedances, star-triangle transformation, Helmholtz-Thevenin's theorem, Helmholtz-Norton's theorem, mesh current analysis, nodal voltage analysis, real power, reactive power, apparent power, power factor, Boucherot's theorem.
• Three-phase systems: three-phase generator, phase and phase sequence, three-phase loads, single line and line-to- line magnitudes, balanced three-phase systems, analysis by reduction to single-phase systems, circuits balanced in voltage, power factor correction, three-phase power (real, reactive and apparent), single-phase wattmeter and varmeter.

### B. Real values and practical problems

Relevance of theoretical topics included in the course must be supported by practical meaning. For instance: linear cir- cuits, usual components (resistor, capacitor, coil), alternate cur- rent, three-phase systems, lumped versus distributed modeling, frequency response, and so on.

With the proposed approach, students will learn typical values for real systems of common use, with the objective to answer questions of interest on those systems just by approximated calculus (find bounds). For instance, values for: wire sizes (typical diameters, lengths, impedance), typical consumptions of domestic appliances (what is the physical operation involved that is consuming the most of the energy in the appliance), typical consumption of some industrial devices. They will learn also how the energy is related to the $CO_2$ emissions, depending on the generation source, and how these values are related to the thresholds for a sustainable development.

Also they will learn how to model some physical systems: PV panels, heat flow in some devices, electromagnetic fields, using equivalent circuits. And how to interpreted the basic components (resistor, capacitor, coil), not only in terms on the relation between current and voltage (typical presentation), but also in terms of the electromagnetic field to model the coupling between components.

They will realize that the sensitivity analysis is a useful tool in practical problems, where the magnitude of many components can take different values, such as loads, or the voltage and frequency output from a generator.

### C. Experiments and measures

Experiments are a very important part in the strategy to provide practical meaning to the theory. Here we consider the following activities related to the experiments:

• Guided experiments in the laboratory to reinforce the basic concepts (Ohm's law, Thevenin's Theorem, power factor). In these laboratory sessions the complete experimental setting is provided: elements to connect, how to connect the elements, what to measure and how to perform the measures.
• Proposed experiments that can be performed at home with low cost components. For instance: Tellegen's Theorem, maximum power transfer, identification of parts of cir- cuits.
• Fitting a model (equivalent circuit) for some system, for which we can measure only some variables and in certain parts of the system, since not all the parts are accessible for measurement. For instance, fitting the equivalent circuit for a battery. In this case some possible equivalent circuits (topologies and type of components in the topology) are provided, and also the battery data-sheet and its response (charging/discharging) under certain con- figurations is provided. Students have to calculate the best fit for the values of the equivalent circuit components, interpret the physical meaning of those components, and decide which is the best equivalent circuit.

TABLE I
## Context for Problems

| Topic | References |
|---|---|
| Air ionizer | [6] (Fig. 4), [7] (Fig. 1) |
| Analog to digital conversión | [8] (Figs. 1 to 5) |
| Battery charger | [9] (Fig. 4) |
| Battery I-V model | [10] (Figs. 1 and 3) [11] (Figs. 1, 2, and 4) |
| Battery lead-acid | [12] (Figs. 1, 2, 4, and 5) |
| Battery Li-Ion | [13] (Figs. 3 and 7) |
| Battery state of charge | [14] (Fig. 1) |
| Battery supercapacitors | [15] (Figs. 2, 4, and 5), [16] (Figs. 2, 3, and 4), [17] (Figs. 2, 3, and 4), [18] (Fig. 1), [19] (Figs. 1 and 2), [20] (Figs. 1, 2, 6, and 10) |
| Bus coupling noise model | [21] (Figs. 6 and 9) |
| Capacitor carbon based | [22] (Fig. 7) |
| Chua's circuit | [23] (Figs. 7 and 8) |
| Class-E power amplifier | [24] (Figs. 3, 6, and 7), [25] (Fig. 1), [26] (Figs. 1, 2, 3, 4, 10, and 11) |
| Coil eddy currents | [27] (Figs. 4 and 5) |
| Contact-less transmission of energy | [28] (Figs. 3, 4, 5, 7, 8, 9, and 10) |
| Converters, PWM small signal | [29] (Figs. 4 and 5) |
| Converters, resonant | [30] (Fig. 4) |
| Defibrillator | [31] (Fig. 7) |
| Electric machine induction | [32] (Fig. 1), [33] (Figs. 2, 7, and 9) |
| Electric machine synchronous | [34] (Figs. 4, 6, and 8) |
| Electrical Muscle Stimulation | [35] (Figs. 1 and 7), [36] (Figs. 2, 3, and 4) |
| Electro acupuncture | [37] (Fig. 3) |
| Electrochemical Impedance | [38] (Fig. 1), [39] (Table II), [40] (Fig. 5 and Tables I-V) |
| Energy harvesting | [41] (Figs. 1 and 8), [42] (Figs. 2 and 3), [43] (Fig. 24), [44] (Figs. 1, 2, 6-12) |
| Ferrite small-signal model | [45] (Fig. 1) |
| Fluorescent Lamps | [46] (Fig. 9), [47] (Fig. 1, ballast), [48] (Figs. 1, 3, 5, 6 and 10) |
| Foster's reactance theorem | [49] (Figs. 1 and 2) |
| Fuel Cell | [50] (Figs. 2, 3, and 4), [51] (Figs. 5 and 12) |
| Gysel Power Divider | [52] (Figs. 1 and 2), [53] (Figs. 1 to 4), [54] (Figs. 1 to 3), [55] (Figs. 1 to 4) |
| Hearing-aid | [56], [57] (Fig. 6), [58] (Figs. 6 and 7) |
| High Voltage Direct Current (HVDC) line | [59] (Figs. 4-10, 18, and 19) |
| Impedance functions in terms of resonant frequencies | [60] (Figs. 1-9) |
| Induction cooker | [61] (Figs. 1, 2, and 6) |
| Ionic polymer-metal composite (IPMC) | [62] (Fig. 6 and Table I) |
| Kron reduction applied to electrical networks | [63] (Figs. 1, 5, 6, 7, and 8) |
| Laser diode | [64] (Fig. 1) |
| Laser effect of diffusion on damping | [65] (Figs. 3 and 5, and Tables I and II) |
| Laser InGaAS | [66] (Fig. 3) |
| Laser N2 | [67] (Figs. 6 to 9) |
| Loudspeaker | [68] (Figs. 3), [69] (Figs. 2, 4, 8, and 9), [70] (Figs. 1 and 2), [71] (Figs. 1, 4, 7-10) |
| Maxwell equations equivalent circuit | [72] (Figs. 1 to 11) |
| Pacemaker wireless charging | [73] (Figs. 2 and 3) |
| Photosynthetic Power Cell | [74] (Fig. 4) |
| PV inverter harmonic distortion | [75] (Figs. 4 and 7) |
| PV Maximum Power Point Tracking (MPPT) | [76] (Fig. 6), [77] (Fig. 6), [78] (Fig. 1) |
| PV panel | [79] (Fig. 1), [80] (Figs. 4, 22 and 23), [81] (Figs. 1, 9, and 10) |
| PV panel 3-D model | [82] (Fig. 1) |
| PV shadowing | [83] (Fig. 1), [84] (Figs. 4 and 7) |
| PV transformerless inverter | [85] (Figs. 8, 10, and 12) |
| RC circuit for one-directional heat flow | [86] (Fig. 1) |
| StatCom | [87] (Fig. 2), [88] (Fig. 2), [89] (Fig. 4) |
| Transformer for integrated circuits | [90] (Figs. 1, 3-6, 8, 9, 12, 13, and 16), [91] (Figs. 1, 6, and 22. and Table V) |
| Transistor, MOSFET thermal circuit | [92] (Fig. 11), [93] (Fig. 2) |
| Transistors, bipolar and MOS | [94] (Figs. 4 and 5) |
| Voltage multiplier | [95] (Figs. 1-4) |

- Using simulation tools, as [96] that allows to simulate circuits using virtual reality, or [97] and [98] that allow to simulate electric and electronic circuits and include guided tutorials. Also there are some games that includes solving electric and electronic circuits as a part, as for instance [99] that mainly is a shooter game.

Experiments allow the students to gain confidence on:

- How the theory represents the real systems. The accuracy of the models and the errors in the measures, and how to manage all that to give proper solutions to practical problems.
- How to use the experiments (measures) to get the data to solve problems in practical situations. In many real problems the data are not provided, and we have to take measures on the real system to obtain them.
- How to use experimental data to build approximate mod- els, as equivalent circuits, by trying different topologies and fitting the values of the circuit components.
- How to use simulation tools to get results and insights on how the circuits are working.

### D. References for self-learning

The implicit assumption in which the proposed framework is based is that students are interested in some practical applications they can find useful in some way, as for instance in their daily life, for future professional development, or just for curiosity. The proposed framework provides some of those practical applications. One of the learning outcomes is to be able to find the proper bibliographic sources to answer the questions. For this reason references as those in Table I are provided to give a starting point.

## V. ROLE-PLAYING GAMES FOR FIRST COURSE ON CIRCUIT ANALYSIS

In this section several role-playing games are drafted using the problems and topics described in the previous sections. Each game consists of several levels, that must be solved sequentially, starting from level 1. In each level the player must solve a certain number of problems before advancing to the next level. Each level contains many more problems than the number needed to pass the level so that the student can choose among those problems. In what follows the game scoring and how it is related to the course assessment is described first, then the contents of four games are presented.

### A. Assessment and Game Scoring

Each problem, according to its difficulty, is associated with a number of "experience points" and "coins", and this information is available for the students. The experience points are used to assess the progress in the course content, and the coins are mainly related to the number of problems solved (training with the same type of problems).

Experience points adds to the "player reputation" and its effect on the reputation value depends on the player current level. The higher the player level the higher the number of experience points needed to increase reputation. Reputation is necessary to unblock levels and some problems inside the levels.

On the other hand, coins are mainly related to the number of problems solved. The number of coins depend linearly on the number of problems and the problems difficulty. They allow to define a mechanism for training in each level (number of problems solved in each level). To go to the next level a certain number of coins is required

Also, the coins can be interpreted as the player incomes in the game, and can be related with the playing time. In order to follow the course scheduling, we set thresholds for the number of coins each player need to reach on certain dates (dates are fixed by the course calendar). In the game it is interpreted as the player need certain incomes to remain in the business.

### B. Repairing Domestic Appliances

In this game the students adopt the role of a technician in an appliance repair shop. There is a sequence of steps to check the device. The appliance is made of a set of parts, and each part have a number of typical failure modes. The typical failure modes considered in this game are: i) short-circuit, ii) open circuit, iii) bypass because of a direct contact or due to other components as for instance powder, and iv) some component is broken: resistor, capacitor, coil, operational amplifier. The failures of some parts usually affect the whole system. Then, we have to start by checking the failures affecting many parts and move towards the failures affecting just single parts. "If the system would be working properly, the measure at this point should have this *value*, but instead, we observe this *value*.". The levels in this game are:

1) Level 1: Initial problems related to switches, and con- nection wires (resistors) in DC. Sometimes the answer is that the switch or the plug are broken. The answer is not just a number they got replacing some values in a formula. They get numbers but they have to interpret those numbers to reach the answer.

- Data: complete information on the circuit topology and the numerical values for all the components (all resistors), and complete information on the tests performed and the values measured. Value of the impedance of the loop to ground.
- Questions: voltage and power for each component in the circuit. Voltage and power if we change the connection topology (for instance from series connection to parallel connection and vice versa). Add the loop to ground and determine in what point is the contact.

2) Level 2: Identify different parts in an equivalent circuit (the circuit can contain loops), using provided measures. Effect of changing the circuit topology.

- Data: the appliance is composed of more than one part, and several equivalent circuits are provided for each part

(only one is right). Some measures are provided as well.

- Questions: Check each part, and determine which is the right equivalent circuit for each part, and what is the problem with the appliance. Also study how the power consumption changes in different parts if the connection topology is changed (series to parallel).

3) Level 3: Repeat the problems in Level 2, but this time considering alternate current (single-phase current) and additional components (coils and capacitors) in the equivalent circuits. In this case circuits are modeled considering impedances and phasors.

4) Level 4: Estimating the circuit topology based on verbal data and frequency response.

- Data: Single-phase current, a verbal description of the appliance, some measures under a number of operation configurations, including some responses to changes in the frequency.
- Questions: Determine the equivalent circuit topol- ogy and estimate the magnitude of the components.

5) Level 5: Appliances with a three-phase circuit.

- Data: Circuit topology, and values of some compo- nents, measures (voltage, current) in some parts of the circuits.
- Questions: Identify the failure and propose a correc- tion. Study the effect (power consumption, power factor, currents, voltages) of changing the topology of a part of the circuit, for instance from delta to star or star to delta, for instance in air-condition machines to get a larger power. Study the advantage of introducing components to improve the power factor.

### C. Electric Circuits in a Conventional Thermal Power Plant

Conventional thermal power plants contain many electric circuits that offers a quite interesting context to pose the problems in an introductory circuit analysis course. Relevant questions on the design and operation of electric circuits in a power plant are related to its security, reliability and economy. Proposed problems are related to the course contents and scheduling following a sequence of five levels:

1) Level 1: At this level only resistive circuits and DC are considered.

- Data: blueprints of the generation plant that contain information on the loads and generators values (only real power) and placement, the connection point to the external grid, and the regions where the cables can be placed, with information on the type of installation (buried or aerial) that can be performed and their cost.
- Questions: Determine the cable routing, testing dif- ferent topologies (a single circuit or several circuits), choosing the wire size among commercial cables, choosing whether to make a buried or aerial installa- tion, calculate voltage drops and power losses in the cables for a number of system load configurations.

2) Level 2: Still considering only resistive circuits and DC. The focus in this level is on reliability and security of the electric circuits.

- Data: Designs from the previous level. Available configurations for the connection to ground circuit, their cost and impedance.
- Questions: Compute short circuit currents for a number of typical faults in the system. Select where to install the protections (overcurrent and overvolt- age), by testing a number of different configurations. With the objective to maximize plant's reliability, by isolating only the smallest possible part of the circuit, while the other parts continue working. Protection against indirect contacts, considering the impedance of the ground connection and determin- ing the threshold of the bypass current to activate the differential protection.

Secondary quest: electric circuits to protect some components (tubes, heat exchangers, metal structures) against corrosion.

3) Level 3: Considering AC and impedances (real and reactive power, and circuits that can contain coils and capacitors). Recheck the questions in previous levels, cable routing and protections, taking into account:

- Real wires are not purely resistive, and have a capacitive and inductive part that can be represented by a impedance.
- Inductive and capacitive coupling between wires placed in the same path (or that are quite close).
- Loads' power factor.

4) Level 4: In this level the focus is placed on the main generator and the transformers considering three-phase circuits:

- Transformation ratio: Calculation of the optimal transformation ratio, taking into account the voltage at the connection point to the external grid, and losses in the transformer and in other circuits of the plant. A simple equivalent circuit for the transformer is considered, the circuit and the values of its components are provided as data.
- Losses in the generator. A simple equivalent cir- cuit is considered for the generator (a synchronous machine). Studying how the equivalent circuit pa- rameters change to provide different values of the power output (voltage and frequency must be in a narrow range), and how that changes are affecting other circuits.
- Problems involving a sensitivity analysis of the system respect to the generator's frequency, the generator's voltage, or the transformer's voltage, are proposed with a particular focus on losses, real power, and reactive power.

5) Level 5: Problems on this level are related to power factor compensation, and analysis of faults in the three- phase circuits. The problems include studying the conve- nience of duplicating circuits and/or creating additional auxiliary connections that can be used in case they are needed by just activating some switches. The objective is to increase the reliability and security in the operation. Calculation of sensitivity factors to quantify how the operation in one branch is affecting other branches.

### D. Electric circuits for households

Households offer a very interesting context for the study of electric circuits, from basic circuits to supply the usual loads at home, to circuits for smart home applications, integration of PV generation (maybe with batteries), or electric vehicle charging.

1) Level 1: First approach considering only resistive circuits and DC.

   - Data: Technical drawing for different households. Data for the typical loads and their range of power consumption. Available sizes for wires, and their cost.
   - Questions: Wire routing, taking into account voltage drops and power losses. Partition in sub-circuits. Approximate cost.

2) Level 2: Still considering only resistive circuits and DC, the focus in this level is on the circuit protections: overcurrent, overvoltage and differential protection.

   - Data: Same data as for Level 1, adding data for the loop to ground.
   - Questions: Where to place the protections, and a calculation of their threshold values. Calculation of the short-circuit currents and the bypass currents for different branches in the circuits.

3) Level 3: Questions in Levels 1 and 2 are recalculated considering single-phase alternating current and circuits with impedances.

   - Data: Complex impedances for wires and loads. Real and reactive power ranges for loads, and their power factor.
   - Questions: Wire routing, protections placement and threshold, voltage drops, power losses (real and reactive power). Calculation of Thevenin equivalent of part of the system when different loads are connected to the circuits.

4) Level 4: A PV system with batteries and a slow charging point for Electric Vehicles (EVs) are added to the house circuits used in Level 3.

   - Data: The same as for Level 3. Data for the PV system. A model (equivalent circuit) for the batteries. An equivalent circuit for the electric vehicle charging point. Some loads, as the charging point for EVs and air-conditioning, that can be connected to a three-phase system.
   - Questions: Size of the PV system and the batteries. Calculation of voltage drops, losses, and power balance when different loads are connected to the system. Sensitivity analysis respect to the voltage and frequency in the main grid. Calculation of components to add for power factor improvement.

5) Level 5: In this level a neighborhood composed of a number of households of the type defined in Level 4 is considered. It is assumed the neighborhood is fed by a three-phase transformer.

   - Data: The same as for Level 4. A map of the neighborhood, describing the placement of the households, the transformer, the street lighting, and the wires routing. A model for the three-phase trans-former. The available wires (sizes, impedances, cost).
   - Questions: Size of the wires to connect the trans-former to the households. Which households con-nect to each phase. Thevenin equivalent for each household, and a sensitivity analysis of its param-eters depending on the loads and sources (PV) connected to the house. Voltage drops and losses in the connections and the transformer for different load configurations, most of them unbalanced. Size of the wires connecting the transformer to the street lighting, voltage drops and power losses in the circuit, protections for the street lighting circuit.

## E. Spaceships and a colony in Mars

In the context of outer space, we have to design the elec-tric circuits to work properly in extreme ambient conditions (temperature, powder). Furthermore, it is of great importance to maximize the circuits' efficiency and to optimize the avail-able generation resources. Since systems of vital importance (supply of air) depend on the power supply, the reliability of the electric system is a very important issue, and the ability to keep the systems working even if some parts fail. Also the question of doing everything at a reasonable cost is important.

1) Level 1: Designing the electric circuits of a small space-ship I. Firstly, it is only considered resistive components, real power and DC current.

   - Data: Technical drawings of the spaceship, includ-ing the placement of the loads, and the regions where the wires can be placed. The value of the loads, and the range of temperatures in different areas. The normalized series of wires that can be used for the circuits. An equivalent circuit for the batteries that can be used in the system.
   - Questions: Decide the batteries location, the wires section, the connection topology and the wires rout-ing to supply the loads, while minimizing the losses and maximizing reliability. For instance, if a battery fails, or if other components fail, it must be possible to isolate that part of the circuit and keep the rest working.

2) Level 2: Designing the electric circuits of a small spaceship II. Still considering only resistive components, real power and DC current. This level is focused on the protections for overcurrent and the addition of PV panels to the system.

   - Data: The data provided in Level 1, and a model of the PV panels including and equivalent circuit for the power point tracking.
   - Questions: Study where to place the protections for overcurrents to maximize circuits reliability, avoid-ing batteries' discharge because of short-circuits. Study the values of the parameters of the power point tracking circuit to maximize the PV panels power output.

3) Level 3: The spaceship from Level 2 arrives to Mars, and we have to design the electric circuits for a small colony there. A

three-phase electric network will be implemented in the colony (some sub-circuits can be in DC).

- Data: A map of the colony with the buildings, the loads, the generator (a small nuclear generator with a three-phase synchronous generator), and the areas that can be used to place the wires. The available sizes for wires. The values of the loads (power factor, real power, reactive power), and the ranges that these values can take. The generator rated capacity, and the curve providing the link between the real and reactive power generated (or an equivalent circuit for the generator).
- Questions: Wires' routing for each phase taking into account voltage drops and losses (real and reactive power) for a number of load configurations (different consumption level and different power factors). Batteries' placement, considering that the batteries operate as loads (charging), or as power sources (discharging).

4) Level 4: The focus in this level is placed on the protection elements for the three-phase installation in the colony, and how to maximize reliability.

- Data: The same data as for Level 3, adding infor- mation on the loop to ground (impedance).
- Questions: Decide the placement of overcurrent, overvoltage and differential protections, to max- imize reliability. Calculate the threshold for the activation of the differential protection (taking into account the loop to ground impedance). Study the circuits for different typical failures (short-circuit, phase-phase, and phase-ground).

5) Level 5: Improving the reliability and efficiency in the colony power supply.

- Data: Some systems, as the system for air cleaning, water recycling and room heating, are indispensable for survival at the colony. Models for PV panels and the power point tracking system.

- Questions: Include some auxiliary (secondary) cir- cuits to increase the system reliability, in such a way that if a wire fails some switches can be activated to isolate that part and to reconnect the load to keep the system working. Improve the loads power factor adding the proper elements for compensation, reducing the losses and voltage drops (a number of load configurations must be taken into account). Study the integration of an additional power source in the system, such as PV panels. Determine how many PV panels, how to connect them, how to tune the power point tracking circuit.

## VI. Conclusions

To summarize, this paper provides examples of problems set in real and practical contexts (situated learning) suitable for implementing PBL in an introductory course on circuit analysis. The associated learning outcomes and the way these problems can be used to build single-player role-playing games are discussed. Additionally, the importance of the assessment process is discussed and the close relation between the achievement of the course learning goals and the game progress is confirmed.